\documentclass{article}
\usepackage[margin=1in]{geometry}
\PassOptionsToPackage{numbers, compress}{natbib}
\usepackage[]{natbib}

\usepackage[utf8]{inputenc} 
\usepackage[T1]{fontenc}    
\usepackage{amsmath}
\usepackage{amsfonts}

\usepackage{graphicx}
\usepackage{subcaption}

\usepackage{amsmath}
\usepackage{amsfonts} 

\usepackage{array}

\title{Learned Proximal Operator for\\ Solving Seismic Deconvolution Problem}
\author{Peimeng Guan$^1$\thanks{Correspondence: \texttt{guanpeimeng@gatech.edu}} , Naveed Iqbal$^2$, Mark A. Davenport$^1$, Mudassir Masood$^2$ \\
$^1$Electrical and Computer Engineering, Georgia Institute of Technology, \\Atlanta, GA 30332, United States\\
$^2$Department of Electrical Engineering, King Fahd University of Petroleum and Minerals, \\Dhahran, Saudi Arabia}

\begin{document}
\graphicspath{{Figs/}}

\date{\vspace{-5ex}}

\maketitle

\begin{abstract}
Seismic deconvolution is an essential step in seismic data processing that aims to extract layer information from noisy observed traces. In general, this is an ill-posed problem with non-unique solutions. Due to the sparse nature of the reflectivity sequence, spike-promoting regularizers such as the $\ell_1$-norm are frequently used. They either require rigorous coefficient tuning or strong assumptions about reflectivity, such as assuming reflectivity as sparse signals with known sparsity levels and zero-mean Gaussian noise with known noise levels. To overcome the limitations of traditional regularizers, learning-based regularizers are proposed in the recent past. This paper proposes a Learned Proximal operator for Seismic Deconvolution (LP4SD), which leverages a neural network to learn the proximal operator of a regularizer. LP4SD is trained in a loop unrolled manner and is capable of learning complicated structures from the training data.
It is worth mentioning that the network is trained with synthetic data and evaluated on both synthetic and real data. LP4SD is shown to generate better reconstruction results in terms of three different metrics as compared to learning a direct inverse.
\end{abstract}

\section{Introduction}
Extracting subsurface layer reflectivities from acquired data is a key goal in seismic data processing. In a geophysical survey, a source wave is generated on the earth's surface using a vibroseis truck and the reflected signals are collected by geophones after the wave reflects back from the boundaries of the layers of the earth. The collected data is known as the \textit{seismic trace} or \textit{trace}. Reflectivity measures the ratio of impedance changes between neighboring earth layers, and the reflectivity series is a list of coefficients collected at varying depths in the earth. This list is ideally a series consisting mostly of zeros, except for depth locations at the layer boundaries. Therefore, the reflectivity coefficients are modeled as a \textit{sparse} signal, with the number of non-zero values denoted as the sequence's \textit{sparsity} \cite{mousa2011processing}. Depending on the layer depth and the sampling frequency, the sparsity level might be different from case to case and is usually unknown in advance. This paper is concerned with estimating the reflectivity series through a process known as deconvolution. 

In attempting to estimate the reflectivity sequence, the source is modeled as a wavelet, and the trace is modeled as the convolution of the source wavelet with the reflectivity sequence \cite{mousa2011processing, oldenburg1983recovery}. Thus reflectivity sequence can be estimated through the deconvolution process. Deconvolution is an important step in the exploration of seismology \cite{van2008robust,bostock1997deconvolution}, as it increases the vertical/depth resolution. However, deconvolution is an ill-posed problem, with non-unique solutions. This behavior stems from the fact that 1) the noise in the measurement process is usually unknown, and 2) the wavelet acts as a band-pass filter to the reflectivity sequence thus losing important high-frequency components in observed traces. Therefore, the peaks in an observed trace might come from a superposition of multiple closely-spaced boundaries. The goal of this paper is to solve the seismic trace deconvolution problem using an amalgamation of optimization- and data-driven methods.

\begin{figure}[t]
    \centering
    \includegraphics[width=0.6\columnwidth]{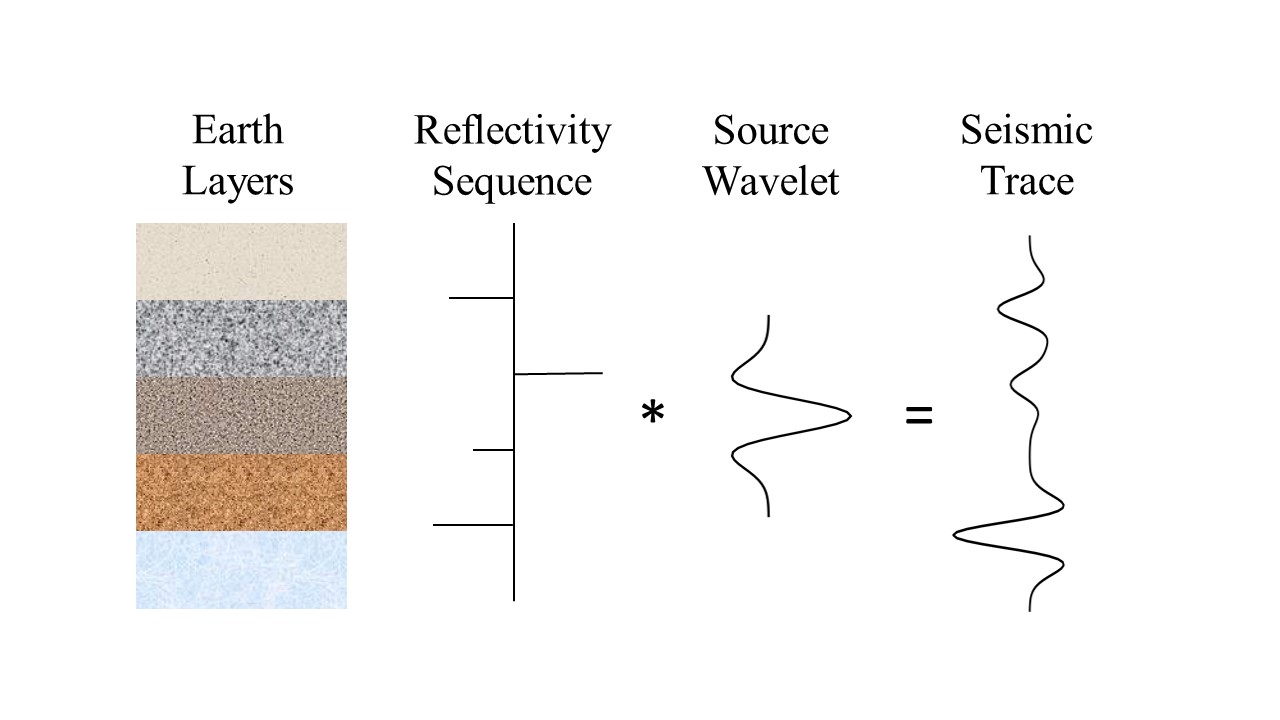}
    \caption{Convolutional model of seismic deconvolution}
    \label{fig:conv_model}
\end{figure}

\subsection{Previous Works}
Classical optimization approaches like Iterative Soft-thresholding Algorithm (ISTA) \cite{https://doi.org/10.48550/arxiv.math/0307152} and Fast Soft-thresholding Algorithm (FISTA) \cite{doi:10.1137/080716542} have been used for reflectivity estimation. These algorithms update the estimate of reflectivity by iteratively applying a proximal operator for $\ell_1$ regularization with a specific sparsity level. This always requires knowing the sparsity level in advance or parameter tuning, and it usually takes hundreds of iterations until the optimization problem converges. By knowing the sparsity level, ISTA can find the location of reflectivities, however, inaccurately reflects the magnitude.

Machine learning has shown its success in learning unknown functions from training data. Works like \cite{russell2019machine, 9431226, PEREG2020103979, phan2021seismic} use neural networks to learn the direct inverse that maps from the observed trace to the reflectivity sequence by knowing that the reflectivity is sparse. Sparse-promoting neural networks \cite{doi:10.1190/geo2020-0342.1} were also developed in this context and can be applied to data with multiple traces. In particular, U-Net promotes sparsity by finding a compressed representation of traces. However, these methods do not use the physical forward operators, but learn a black-box inverse mapping, which is lack of interpretability. In addition, training a robust network to unseen sparsity requires a large amount of data with various sparsity levels.

To increase interpretability, Loop Unrolling (LU) algorithms \cite{9363511} unfold the optimization steps into a sequence of weight-tying neural networks. 
Such algorithms have shown success in wide applications because they utilize the forward operator as part of the recovery process \cite{9363511, andrychowicz2016learning}. Some of these applications include medical imaging reconstruction \cite{https://doi.org/10.48550/arxiv.2102.07944, liang2019deep, putzky2019invert}, image deblurring \cite{https://doi.org/10.48550/arxiv.2102.07944, li2019deep, 8950351}, compressive sensing \cite{https://doi.org/10.48550/arxiv.2102.07944, yang2018admm, li2020end}, etc. Many other variations of LU algorithms use more advanced network architectures for learned regularizers that make the overall network more powerful. For example, a transformer is trained in \cite{https://doi.org/10.48550/arxiv.2203.08213} as the regularizer to improve the performance for Magnetic Resonance Imaging reconstruction, and \cite{https://doi.org/10.48550/arxiv.2102.07944} extends the LU iterations to a potentially infinite number of layers until converging to a fixed point solution, so one can control the output quality based on needs in the evaluation stage. Many of them achieve state-of-the-art results in different tasks.

The authors in \cite{gregor2010learning, zhang2018ista, xiang2021fista} adopt the idea of LU for sparse recovery problems with specific designs for the proximal operators. Considering the reflectivity sequence as a sparse signal, \cite{https://doi.org/10.48550/arxiv.2104.04704} unfolds the iterative soft-thresholding algorithm and replaces the proximal operator with a minimax-concave penalty to retrieve layer reflectivities. However, this setup limits its direct application on 2D data. This network is trained only on 1D synthetic data, and is evaluated trace-by-trace to obtain 2D recovery. Hence useful information from neighboring traces is not utilized. In addition, similar to ISTA and FISTA, \cite{https://doi.org/10.48550/arxiv.2104.04704} recognizes the location of non-zeros but inaccurately retrieves the magnitudes due to the restricted network setup for the proximal operator.

\subsection{Our Contribution}
This work proposes a novel approach named Learned Proximal operator for Seismic Deconvolution (LP4SD), a loop-unrolled architecture for learning regularizers in seismic deconvolution. Unlike networks that learn the direct inverse (i.e. U-Net), the proposed network breaks down the task of learning deconvolution into $K$ smaller/easier tasks of learning proximal operators, using the knowledge of the forward operator. Unlike classical optimization approaches that impose potentially erroneous predetermined regularizers in the objective function, LP4SD learns the proximal operator from data to avoid making incorrect assumptions of the reflectivity and tedious hyperparameter tuning. This fully data-driven proximal operator also allows reconstruction of multiple traces simultaneously.
This paper employs LP4SD with a known sampling frequency and source wavelet.  Several constraints are applied to stabilize the process: 1) the trainable step-size in loop unrolled iterations is wrapped in a range function to avoid explicit hyper-parameter tuning and to ensure stable convergence, 2) the measurement is passed as a direct input to the network together with intermediate reconstruction to efficiently correct the noise and artifacts. To our knowledge, this is the first paper introducing LU with a generic data-driven regularizer in solving seismic deconvolution problems. In the experiments, the network is trained on synthetic data and evaluated in both synthetic and real data. We show that LP4SD outperforms the sparsity-promoting U-Net with less number of layers in both single trace and multiple traces cases, and is more robust to unknown sparsity and noise levels.\\

\begin{figure}[t]
    \centering
    \includegraphics[width=0.4\columnwidth]{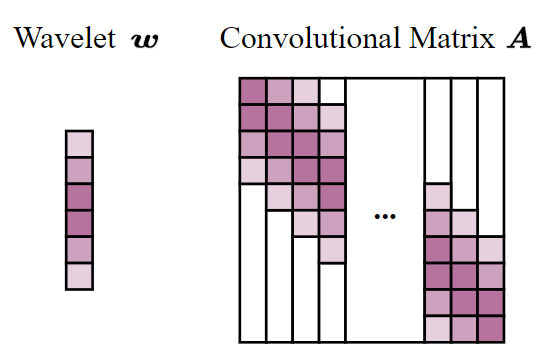}
    \caption{Generating convolutional matrix $\boldsymbol{A}$ from wavelet $\boldsymbol{w}$, shaded entries represent nonzero values.}
    \label{fig:conv_mtx}
\end{figure}

The rest of the paper is organized as follows. Section \ref{sec:description} states the seismic deconvolution problem and its mathematical model. Section \ref{sec:LU-Porx} overviews the optimization program for recovering reflectivity, then introduces the optimization-guided neural network and how to train this network. Section \ref{sec:exp} presents the experiment results on synthetic data, baseline model (Marmousi2), and real data. Finally, Section \ref{sec:conclusion} concludes the paper.

\section{Problem Description} \label{sec:description}
The objective of seismic deconvolution is to reconstruct the reflectivity sequence from noisy traces. The received trace $\boldsymbol{y} \in \mathbb{R}^{n}$ ($n$ denotes the number of time samples) is modeled as \begin{equation} \label{eq:conv}
    \boldsymbol{y} = \boldsymbol{w} * \boldsymbol{x} + \boldsymbol{\epsilon},
\end{equation}
where $\boldsymbol{w} \in \mathbb{R}^{d}$ represents the source signal wavelet, $\boldsymbol{x} \in \mathbb{R}^{n}$ is the reflectivity sequence and $*$ denotes the convolution operation. Further, $\boldsymbol{\epsilon} \in \mathbb{R}^{n}$ represents the unknown noise. Generally, $d\ll n$ and $\boldsymbol{w}$ are assumed to be known. The convolutional model of \eqref{eq:conv} is illustrated in Fig. \ref{fig:conv_model} for clarity. Note that convolution is a linear operation and thus \eqref{eq:conv} can be written as
\begin{equation} \label{eq:Ax}
    \boldsymbol{y} = \boldsymbol{A}\boldsymbol{x} + \boldsymbol{\epsilon},
\end{equation}

where $\boldsymbol{A} \in \mathbb{R}^{n\times n}$ is the convolutional matrix and follows the Toeplitz structure formed using $\boldsymbol{w}$. The columns of $\boldsymbol{A}$ can be formed sequentially by shifting the vector $\boldsymbol{w}$ by one time-step at a time. Figure \ref{fig:conv_mtx} illustrates this process, where the entries in white are zero-paddings. $\boldsymbol{A}$ is also called the forward operator of the seismic deconvolution problem.

\begin{figure*}[t]
    \centering
    \includegraphics[width=0.9\textwidth]{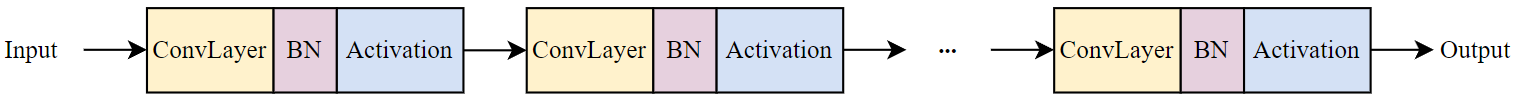}
    \caption{Architecture of a Convolutional Neural Network}
    \label{fig:CNN}
\end{figure*}

An estimate of the reflectivity sequence $\hat{\boldsymbol{x}}$ can be obtained by solving the following optimization problem,
\begin{equation}
    \hat{\boldsymbol{x}} = \min_{\boldsymbol{x}}\frac{1}{2} \|\boldsymbol{y}-A\boldsymbol{x}\|^2_2 + \gamma \, r(\boldsymbol{x}).
    \label{Eq:least-square}
\end{equation}
Here, $\|\boldsymbol{y}-A\boldsymbol{x}\|^2_2$ penalizes the data misfit using current estimate of $\boldsymbol{x}$. Furthermore, $r: \mathbb{R}^n \rightarrow \mathbb{R}_{\geq 0}$ is a regularization function.
The choice of the $r$ depends on the prior beliefs of the underlying signal $\boldsymbol{x}$ and the computational feasibility. For example, $\ell_2$-norm encourages minimum norm solutions, while $\ell_0$-norm encourages sparse solutions \cite{taylor1979deconvolution}. In addition, $\gamma$ is the regularization hyper-parameter that is usually well-tuned to balance the data misfit and the regularization term.

The problem in (\ref{Eq:least-square}) can be solved via iterative optimization methods, such as gradient descent. The gradient descent method iteratively updates the estimate by taking a step in the direction of the negative gradient of the objective function. For  (\ref{Eq:least-square}), this update takes the form
\begin{equation}\label{eq:GD}
    \hat{\boldsymbol{x}}_{k+1} = \hat{\boldsymbol{x}}_k + \eta \boldsymbol{A}^\top (\boldsymbol{y}-\boldsymbol{A}\hat{\boldsymbol{x}}_k) - \eta \nabla r(\hat{\boldsymbol{x}}_k),
\end{equation}
for $k = 1,2,3,...$, where $\eta >0$ represents a constant step-size for all $k$. Note that the step-size can vary for each $k$, but this work only focuses on the constant step-size scenario. While Gradient Descent is an attractive algorithm, it may diverge when the regularization function is non-differentiable. An alternative to overcome this is the proximal gradient method, where in each iteration, the proximal operator of $r$ is applied to the gradient update of the data misfit term, as shown below,
\begin{equation}\label{eq:prox}
    \hat{\boldsymbol{x}}_{k+1} = prox_{\eta,r}(\hat{\boldsymbol{x}}_k + \eta \boldsymbol{A}^\top(\boldsymbol{y}-\boldsymbol{A}\hat{\boldsymbol{x}}_k)).
\end{equation}
The proximal operator enforces the structure that the regularization function $r$ attempts to encourage. For simple regularizers like $\ell_1$ and $\ell_2$, the proximal operators have closed-form solutions but for general regularizers, closed-form solutions may not exist.

Although in seismic deconvolution problems, the reflectivity sequence $\boldsymbol{x}$ is always treated as a sparse signal, its sparsity, determined by the earth structure as well as the sampling frequency, is unknown in advance. In particular, a high sampling frequency of traces is synonymous to acquiring a higher number of samples along the vertical direction. A direct consequence of this is to have a higher number of zeros in the estimated reflectivity and hence a higher sparsity. Similarly, low sampling rates are translated into a less sparse reflectivity. Together with the unknown measurement noise $\boldsymbol{\epsilon}$, we want a general regularizer that can not only capture complicated structures of $\boldsymbol{x}$ but its proximal operator can also be easily solved.

\begin{figure}[b]
    \centering
    \includegraphics[width=0.6\columnwidth]{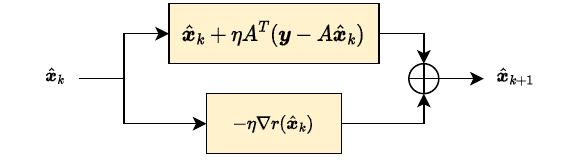}
    \caption{Loop Unrolling Gradient Descent Network, where $-\eta \nabla r(\cdot)$ is replaced with a neural network}%
    \label{fig:LU_grad}%
\end{figure}

\section{Optimization-guided Neural Networks} \label{sec:LU-Porx}
This section will first overview important neural network concepts and introduce the proposed optimization-guided neural network that can (i) relax the prior assumption about the reflectivity sequences and noise, (ii) learn the complicated structures of $\boldsymbol{x}$ from training examples, and (iii) use the knowledge of the forward operator. 

\subsection{Neural Network Overview}
Neural networks can learn complex structures from examples. It is composed of a sequence of linear and non-linear layers with trainable parameters. A loss is calculated between the ground-truth data and the output of the neural network, and the network parameters are updated by calculating the Jacobian loss with respect to the input.

Convolutional Neural Networks (CNN) \cite{lecun1995convolutional} are commonly used in image processing tasks, due to their effectiveness in obtaining local correlation from neighboring pixels. There are many variations of CNN, but a classic architecture contains a sequence of convolutional layers followed by normalization and nonlinear activation layers. In each convolutional layer, a convolutional kernel is applied to compute the spatial correlation between the kernel and the input. Training CNN will update the parameters in kernels. BatchNorm (BN) and GroupNorm (GN) are commonly used as normalization layers, which normalize the output from the previous layer to alleviate the internal covariate shift \cite{ioffe2015batch}, thus allowing stable training of deeper neural networks. Activation layers introduce non-linearity to the neural network, where some commonly used activation functions are ReLU, LeakyReLU, Sigmoid and etc. Figure \ref{fig:CNN} shows the architecture of a simple CNN.

A larger kernel size ($\kappa$) of a convolutional kernel can capture information from a broader range, but it means more parameters to train. Thus the kernel size is usually kept small due to memory constraints. This paper also examines the effect of kernel size in seismic deconvolution tasks. 

\begin{figure}[b]
    \centering
    \includegraphics[width=0.6\columnwidth]{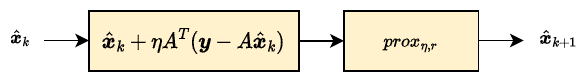}
    \caption{Loop Unrolling Proximal Gradient Descent Network, where $prox(\cdot)$ is replaced with a Neural Network}%
    \label{fig:LU_prox}%
\end{figure}

\begin{figure*}[htp]
    \centering
    \includegraphics[width=0.85\textwidth]{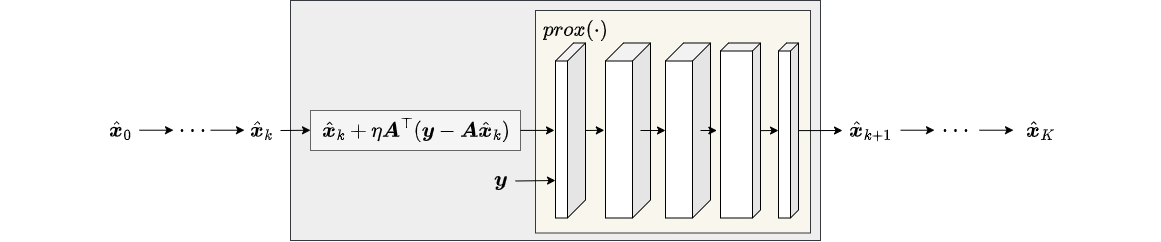}
    \caption{Overall Architecture of LU4SD, where the outer box in gray is an update step at iteration $k$, and the inner box in yellow is the learned proximal operator by a 5-layer CNN. The input to the CNN is a concatenation of $\hat{\boldsymbol{x}}_k + \eta A^\top(\boldsymbol{y}-A\hat{\boldsymbol{x}}_k)$ and $\boldsymbol{y}$.}
    \label{fig:lp4sd_architecture}
\end{figure*}

\subsection{Learned Proximal Operator for Seismic Deconvolution}
Although in classical deconvolution algorithms, the regularizers always use $\ell_0$- and $\ell_1$-norm, they oversimplify the model for reflectivity sequence with the presents of unknown noise and sparsity level.
To avoid simple assumptions on $r$, we can instead learn the regularization update using a neural network. In particular, the gradient update of $r$ in (\ref{eq:GD}) and the proximal operator in (\ref{eq:prox}) can be replaced by neural networks, and the entire optimization iterations in Gradient Descent (\ref{eq:GD}) and Proximal Gradient Descent (\ref{eq:prox}) can be unfolded into a sequence of weight-tying neural network blocks, which refers to Loop Unrolling (LU) \cite{9363511}. One optimization iteration in the networks for gradient descent and proximal gradient descent is illustrated respectively in Figures \ref{fig:LU_grad} and \ref{fig:LU_prox}. Notice that $\hat{\boldsymbol{x}}_k + \eta \boldsymbol{A}^{\top}(\boldsymbol{y}-\boldsymbol{A}\hat{\boldsymbol{x}}_k)$ incorporate the knowledge of the forward operator $\boldsymbol{A}$, but is independent of the regularizer.

This paper only focuses on the proximal gradient method for seismic deconvolution. Because the underlying reflectivity sequence is compressible meaning most reflectivity coefficients are close to zero, $\ell_1$-norm is used as the regularization function. In the gradient descent version of LU, the network tries to learn the gradient of $\ell_1$-norm which is discontinuous. Learning discontinuity is, in general, a hard task for a feedforward neural network such as CNN, hence, the proximal gradient method is the primary focus. 

The gradient of the least-squares term in (\ref{Eq:least-square}) incorporates the forward operator and indicates the update direction to match the noisy trace, and the proximal operator is a correction term that learns complicated structures of $\boldsymbol{x}$. Thus the proximal operator can also be viewed as a denoising process. Different network architectures can serve as a denoiser, but this paper illustrates the idea using only Convolutional Neural Networks. We call it the Learned Proximal operator for Seismic Deconvolution (LP4SD) which is illustrated in Figure \ref{fig:lp4sd_architecture}, where a 5-layer CNN with 64 kernels in hidden layers is used. The detailed layer structures are listed in Appendix. At each loop unrolling iteration $k = 1,..., K$, the estimate $\hat{\boldsymbol{x}}_{k+1}$ is updated as
\begin{equation}
    \hat{\boldsymbol{x}}_{k+1} = CNN(\hat{\boldsymbol{x}}_k + \eta \boldsymbol{A}^\top(\boldsymbol{y}-\boldsymbol{A}\hat{\boldsymbol{x}}_k), \boldsymbol{y}),
\end{equation}
Notice that the measurement $\boldsymbol{y}$ is fed to the CNN to stabilize training together with the gradient step in the data domain. Since LP4SD is trained end-to-end (will discuss more detail in Section \ref{sec:training_process}), there is no explicit constraint on intermediate results $\hat{\boldsymbol{x}}_k$, thus $\boldsymbol{y}$ can efficiently correct the reconstruction errors at the intermediate stages.

\begin{figure*}[hbt]
\centering
\includegraphics[width=\textwidth]{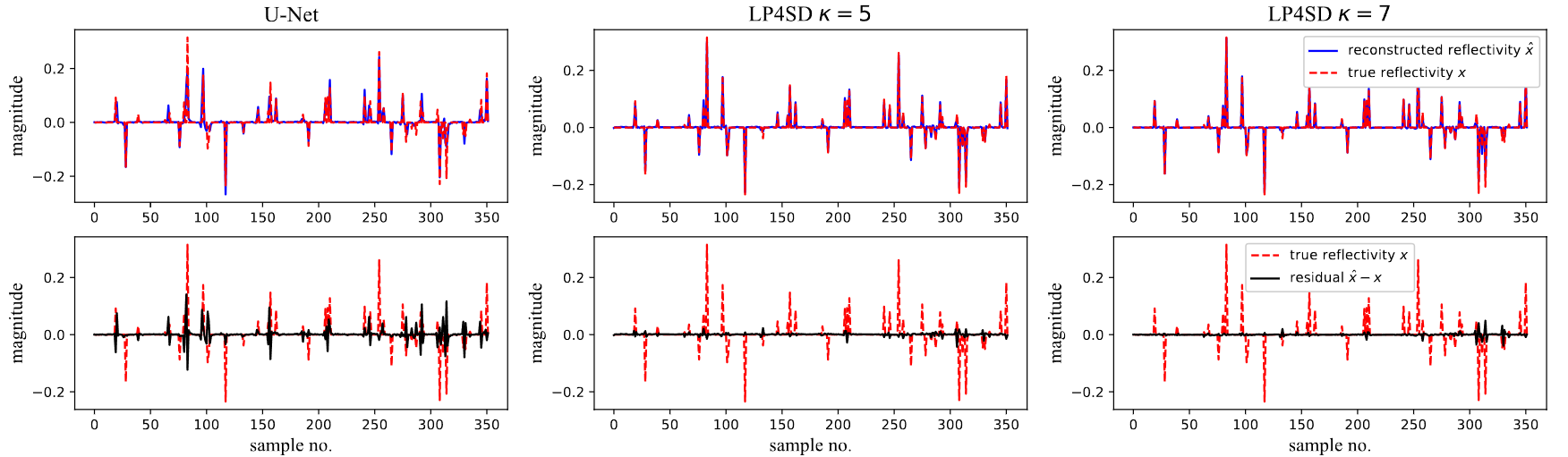}
\caption{2D synthetic reconstruction of selected trace using U-Net, LP4SD with kernel size of 5 and 7. The first row compares the reconstructed reflectivity sequences, $\hat{\boldsymbol{x}}$, and the second row shows the residual, $\hat{\boldsymbol{x}} - \boldsymbol{x}$.}
\label{fig:syn_2D_single_trace}
\end{figure*}

\subsection{Training Process}\label{sec:training_process}
Given the forward operator $\boldsymbol{A}$ and a pair of ground truth reflectivity and corrupted trace $(\boldsymbol{x}, \boldsymbol{y})$, we initialize $\hat{\boldsymbol{x}}_0$ with a signal in the same domain as $\boldsymbol{x}$. Some common choices of $\hat{\boldsymbol{x}}_0$ are $\boldsymbol{A}^\top\boldsymbol{y}$ when $\boldsymbol{x}$ and $\boldsymbol{y}$ are from different domains, and $\boldsymbol{y}$ when they are from the same domain. Then after $K$ loop unrolled iterations, compute the Mean-Squared Error (MSE) between $\hat{\boldsymbol{x}}_K$ and $\boldsymbol{x}$ and backpropagate using the MSE loss, which is referred as end-to-end training.

 Notice that the choice of step-size $\eta$ affects the convergence in the proximal gradient method. In LP4SD, the step-size $\eta$ is registered as a network parameter along training to avoid explicit hyper-parameter tuning for $\eta$, but imposing bounds on $\eta$ is essential. Intuitively, a small step-size results in slow convergence, and a large step-size diverges the process. The range of the step-size that leads to a proper convergence is well-studied for convex optimization \cite{}, which always imposes further assumptions on the objective function, such as strong convexity and smoothness. The range of the step-size always depends on the parameters of those assumptions. In LP4SD, since the proximal operator is an arbitrary non-convex neural network, the classical theorems do not apply anymore, but idea of ensuring a range is still necessary. Therefore, based on the sensitivity analysis, we heuristically assign an initial value to $\eta$ and constrain the network to update $\eta$ within a range of $0$ and $0.15$ by wrapping $\eta$ with a Sigmoid function. So the constrained step-size $s$ becomes
\begin{equation}
    s(\eta) = 0.15 \, \frac{1}{1+e^{-\eta}}.
\end{equation}
It empirically ensures each proximal gradient step updates properly and the input at each loop unrolling iteration to update in a reasonable range.

\section{Experiments and Discussions}  \label{sec:exp}
\begin{table}[b]
    \centering
    \caption{\label{tab:LU_ks5} Architecture of LP4SD $\kappa=5$. When $\kappa=7$, change padding to 3 to match the dimension.}
    \begin{tabular}{ cc } 
         Layer & Details \\
         \hline
         L1 & conv($C_{in}$:2,  $C_{out}$:64, $ks$:5, $s$:1, $pad$:2) + GN + ReLU \\
         L2 & conv($C_{in}$:64, $C_{out}$:64, $ks$:5, $s$:1, $pad$:2) + GN + ReLU \\
         L3 & conv($C_{in}$:64, $C_{out}$:64, $ks$:5, $s$:1, $pad$:2) + GN + ReLU \\
         L4 & conv($C_{in}$:64, $C_{out}$:1, $ks$:5, $s$:1, $pad$:2) + GN + ReLU \\
         L5 & conv($C_{in}$:1, $C_{out}$:1, $ks$:1, $s$:1, $pad$:0) \\
    \end{tabular}
\end{table}

In this work,  the experiments to assess the efficacy of the proposed approach for the estimation of reflectivity sequence in various scenarios are presented. The experiments implement the proposed network with kernel sizes of 5 and 7 in the CNN, denoted as LP4SD ($\kappa=5$) and LP4SD ($\kappa=7$) respectively. Table \ref{tab:LU_ks5} shows the layer details for CNN in LP4SD with a kernel size of 5. We denote $C_{in}$ as the input channels, $C_{out}$ as the output channels, $\kappa$ as the kernel size, $s$ as the stride, and $pad$ as the padding number. GroupNorm (GN) and ReLU activation are appended after each hidden convolutional layer. Padding = 3 when $\kappa=7$ to match the dimension. 

Then, the proposed networks are compared to the sparse-promoting U-Net. U-Net is a deep convolutional network with a narrow ``neck" in the middle, where the input features are transformed onto a lower dimensional space thus helps to promote sparsity of the output. The detailed structure of U-Net is presented in Appendix. 

We consider both single-trace (1D) and multiple-trace (2D) reconstruction scenarios in this work. In 1D, a trace $\boldsymbol{y} \in \mathbb{R}^n$ follows the convolutional model in Eq \eqref{eq:conv}. In 2D, the received trace  $\boldsymbol{Y}\in \mathbb{R}^{n\times m}$ and the reflectivity to reconstruct $\boldsymbol{X} \in \mathbb{R}^{n\times m}$ are both matrices, where $n$ is the number of time samples and $m$ is the number of traces collected. Thus, Eq \eqref{eq:conv} becomes
\begin{equation}
    \boldsymbol{Y} = \boldsymbol{A} \boldsymbol{X} + \boldsymbol{\epsilon}
\end{equation}
and the proximal update rule becomes
\begin{equation}
    \boldsymbol{\hat{X}}_{k+1} = prox_{\eta, r}(\boldsymbol{\hat{X}}_{k} + \eta \boldsymbol{A}^\top (\boldsymbol{Y} - \boldsymbol{A} \boldsymbol{\hat{X}}_{k}))
\end{equation}
The convolutional layers in LP4SD are extended to 2D in multiple-trace scenarios. For example, conv1d($C_{in}$:64, $C_{out}$:64, $\kappa$:7, $s$:1, $pad$:3) becomes conv2d($C_{in}$:64, $C_{out}$:64, $\kappa$:(7,7), $s$:(1,1), $pad$:(3,3)).
Due to the narrow-neck design of the U-Net, for 2D input with higher input dimension, a deeper network is required to narrow the ``neck" and to further promote sparsity. The first three columns in Table \ref{tab:1D_synthetic} summarize the methods to compare in various scenarios.

All networks are trained using synthetic data, then evaluated using the benchmark synthetic Marmousi2 model, and finally evaluated using real data.

\subsection{Evaluation Criteria}
The performance of the proposed method is evaluated using three different metrics. These metrics measure the similarities between the true and estimated reflectivities. The evaluation metrics and their formulas are listed below.\\
\textbf{Mean-squared error (MSE)}
\begin{equation}
    MSE(\hat{\boldsymbol{x}}, \boldsymbol{x}) =
    \|\hat{\boldsymbol{x}}-\boldsymbol{x}\|^2_2,
\end{equation}
\textbf{Correlation coefficient}
\begin{equation}
    \gamma(\hat{\boldsymbol{x}}, \boldsymbol{x}) = \frac{\hat{\boldsymbol{x}}^\top\boldsymbol{x}}{\|\hat{\boldsymbol{x}}\|_2 \|\boldsymbol{x}\|_2},
\end{equation}
\textbf{Reconstruction quality}
\begin{equation}
    Q(\hat{\boldsymbol{x}}, \boldsymbol{x}) = 10 \log_{10}\left(\frac{||\boldsymbol{x}||_2^2}{\left\lVert\boldsymbol{x} - \hat{\boldsymbol{x}}(\hat{\boldsymbol{x}}^\top\boldsymbol{x})  / ||\hat{\boldsymbol{x}}||^2_2 \right\rVert^2_2} \right).
\end{equation}

\subsection{Training on Synthetic Data}
All networks are trained on the same set of synthetic data, which is generated following the procedure in \cite{8641282}. 2D data is generated at $m=352$ traces per shot, collecting $n=352$ data points per trace along the Earth's depth at a sampling frequency of 500 Hz. Notice that $m$ and $n$ do not need to be the same.
We use 40 Hz Ricker Wavelet to generate the observed trace $\boldsymbol{y}$ by following the model in \eqref{eq:conv}. Additive white Gaussian noise corresponding to various signal-to-noise ratios (SNR) is used to generate the data. SNR is defined as
\begin{equation}
    \text{SNR} = 10 \log_{10} \left( \frac{\|\boldsymbol{w}*\boldsymbol{x}\|^2_2}{\|\boldsymbol{\epsilon}\|_2^2}\right),
\end{equation}
where $\|\boldsymbol{w}*\boldsymbol{x}\|^2_2$ is the energy of the clean signal and $\|\boldsymbol{\epsilon}\|_2^2$ is the energy of noise. 1D data are selected from a random trace in a shot from 2D data. In the noiseless case, 8,000 samples are generated for training and 1,000 for testing, for other noise levels, 20,000 samples are generated for training and 2,000 for testing. All traces are normalized $\boldsymbol{y}/\max(|\boldsymbol{y}|)$ such that $\max |\boldsymbol{y}| = 1$. Notice that the mean of $\boldsymbol{y}$ is not subtracted to preserve the magnitude of zeros in reflectivity. The magnification factor, $\max(|\boldsymbol{y}|)$, is recorded to bring the recovered reflectivity to its correct level. The same procedure is applied in the testing phases.

\begin{table}
    \centering
    \caption{\label{tab:1D_synthetic} Synthetic 1D and 2D testing results, where the best performances for each metric are in bold.}
    \begin{tabular}{ cccccc } 
            &  Methods          & \# Layers    & MSE    & $\gamma$ &  Q  \\
         \hline
            & U-Net             & 18        & 0.00174   & 0.926   & 9.12\\
         1D & LP4SD ($\kappa=5$)      & 5         & 0.00122   & 0.948   & 10.67\\
            & LP4SD ($\kappa=7$)      & 5         & \textbf{0.00081}  & \textbf{0.966}  & \textbf{12.38}\\ \hline
            & U-Net             & 22        & 0.000625   & 0.890   & 7.40\\
         2D & LP4SD ($\kappa=5$)      & 5         & 0.000040   & 0.993   & 20.22\\
            & LP4SD ($\kappa=7$)      & 5         & \textbf{0.000037}  & \textbf{0.994}  & \textbf{20.86}\\
    \end{tabular}
\end{table}

\subsection{Testing on Synthetic Data}
\begin{figure}
    \centering
    \includegraphics[width=0.7\columnwidth]{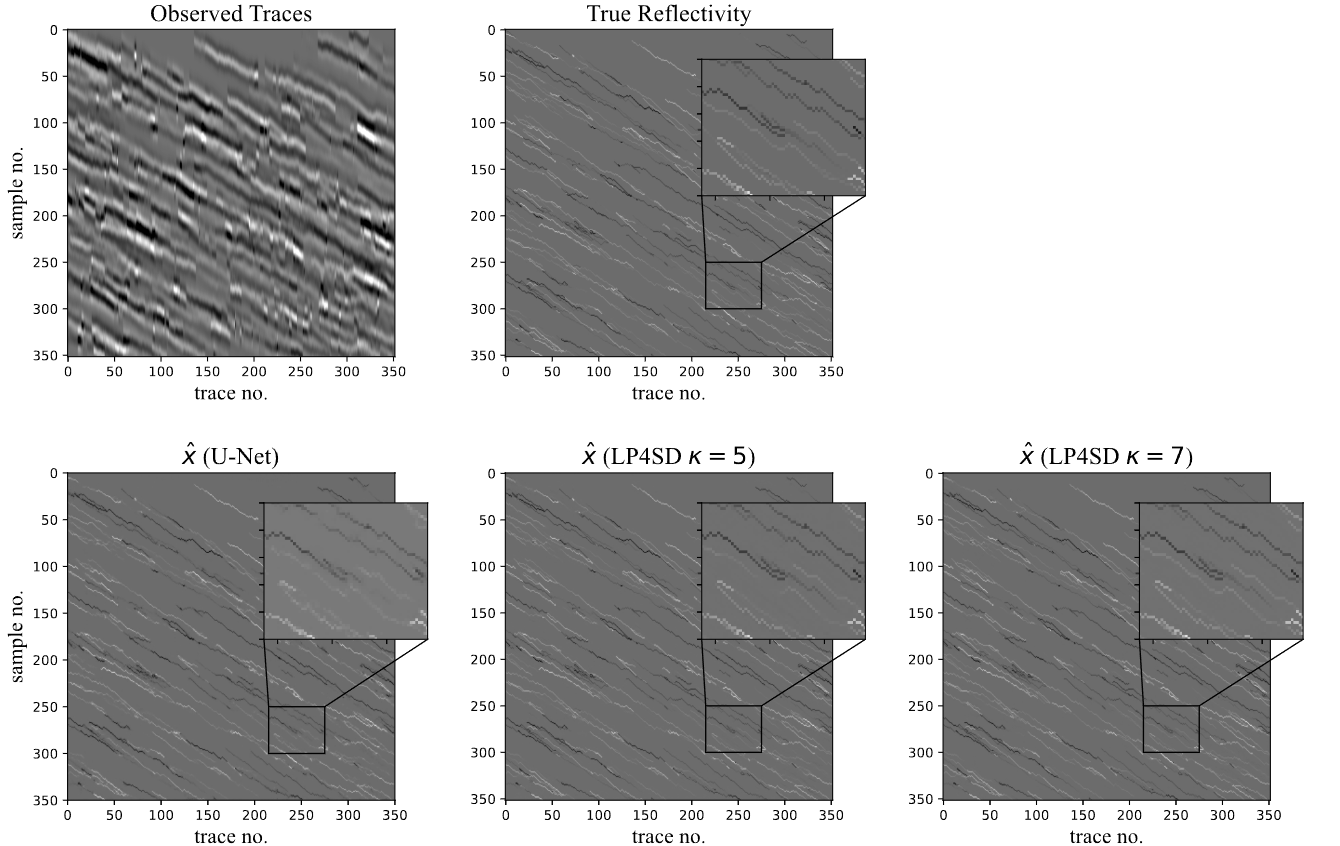}
    \caption{The first row shows the observed traces and the true reflectivity for 2D synthetic data. The bottom row shows the reconstructed reflectivities, where the selected regions are enlarged. Two variants of LP4SD restore more detailed layer structures than U-Net. Results are better viewed electronically.}
    \label{fig:syn_2D}
\end{figure}
\subsubsection{Noiseless Case}
First, the LP4SD methodology is validated for recovering reflectivity in the absence of noise.  Numerical results for 1D and 2D synthetic data are highlighted in Table \ref{tab:1D_synthetic}. In both cases, the LP4SD algorithm outperforms the U-Net with fewer layers. Furthermore, since 2D data contains additional information from neighboring traces, it achieves better reconstruction quality.  In 1D, a larger kernel size improves the quality in all metrics by a noticeable amount, while in 2D, a large kernel size still leads the results in all metrics but the advantage is minimal. A major takeaway is that even without prior assumption to reflectivity $\hat{\boldsymbol{x}}$, LP4SD can recover it more accurately. Figure \ref{fig:syn_2D_single_trace} shows the reconstruction in blue and the reconstruction error (residual) in black of 2D data using U-Net and LP4SD with kernel sizes of 5 and 7. LP4SD significantly reduces the error. Figure \ref{fig:syn_2D} shows the 2D image of the reflectivity sequences, where both variants of LP4SD restore more detailed layer structures, which can be viewed in the enlarged boxes.

\begin{figure}
    \centering
    \includegraphics[width=0.7\columnwidth]{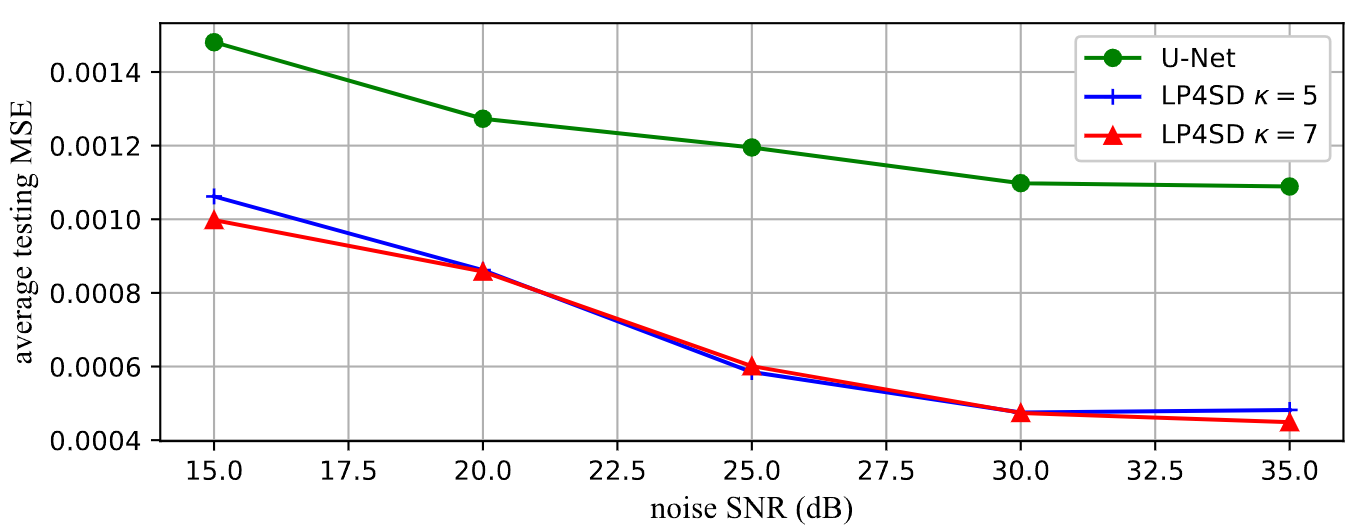}
     \caption{Average 2D testing MSE with additive noise at different SNRs, where LP4SD shows significant MSE reduction at all noise levels.}
    \label{fig:SNR}
\end{figure}
\subsubsection{Noisy Case}
The reconstructions at various noise levels are also analyzed and compared.  The networks are trained using SNRs of 15, 20, 25, 30, and 35 dB. Figure \ref{fig:SNR} illustrates the testing MSE for 2D data. LP4SD outperforms U-Net in all cases and the average MSE gap is significant. This figure also shows that the kernel size in LP4SD is not a determinant factor in 2D reconstruction. In general, CNN with a smaller kernel size runs longer because the convolutional kernels take longer to scan over the input, whereas CNN with a larger kernel size has a tradeoff of more network parameters to train. One has the flexibility of choosing the kernel size according to their needs.

\subsection{Testing on Marmousi2 Model}
\begin{figure}[htp]
\centering
\subfloat[Full samples]{%
  \includegraphics[clip,width=0.8\columnwidth]{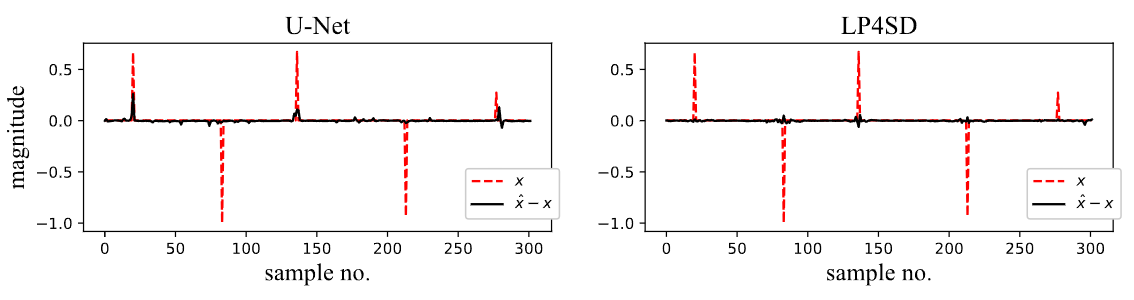}%
}

\subfloat[Downsampled by 2]{%
  \includegraphics[clip,width=0.8\columnwidth]{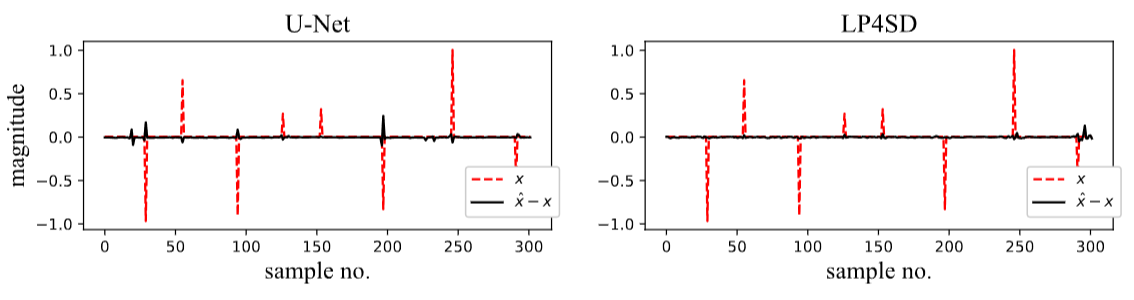}%
}

\subfloat[Downsampled by 3]{%
  \includegraphics[clip,width=0.8\columnwidth]{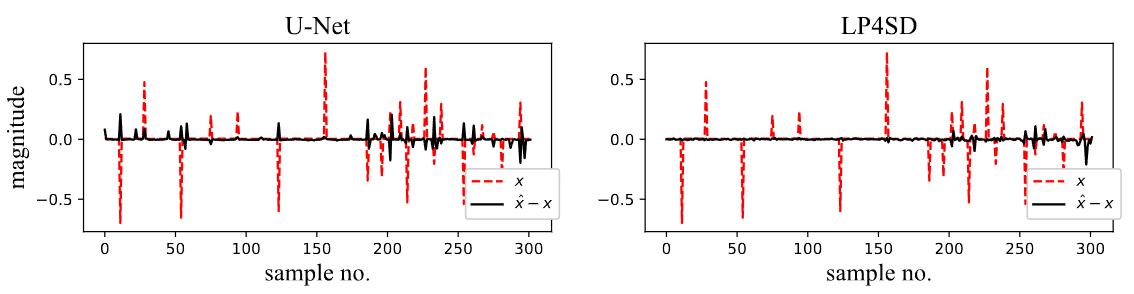}%
}
\caption{Residual of reconstruction, $\hat{\boldsymbol{x}}-\boldsymbol{x}$, on Marmousi2 data (trace 5) using 1D U-Net and LP4SD with kernel size 7. The dotted red line shows the true reflectivity and the solid black line shows the residual.}
\label{fig:1D_res_marmousi}
\end{figure}

Marmousi2 model \cite{martin2006marmousi2} reflects synthetic elastic data, where the survey covered 17 km along the surface and 3.5 km depth. It is a widely-used benchmark dataset for complex seismic processing tasks. The impedance model is obtained from element-wise multiplication between the density and velocity model, and the true reflectivity is then computed from the change of impedance in the vertical direction. The observed trace is obtained by convolving a 40 Hz Ricker wavelet with the reflectivity profile plus random noise. Marmousi2 is a densely sampled model and we also downsample the velocity and density model to resemble more compact reflectivity sequences, as can be seen in the dotted red lines in Figure \ref{fig:1D_res_marmousi}.

\begin{figure}[htp]
\centering
\subfloat[Full samples]{%
  \includegraphics[clip,width=0.8\columnwidth]{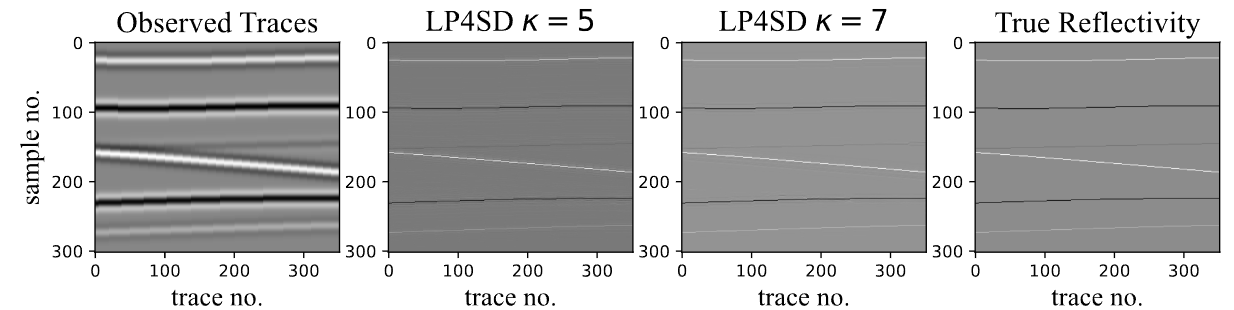}%
}

\subfloat[Downsampled by 2]{%
  \includegraphics[clip,width=0.8\columnwidth]{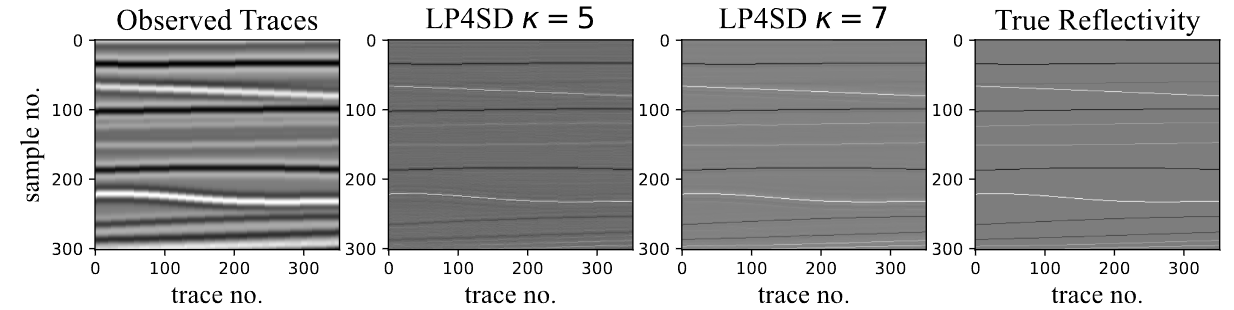}%
}

\subfloat[Downsampled by 3]{%
  \includegraphics[clip,width=0.8\columnwidth]{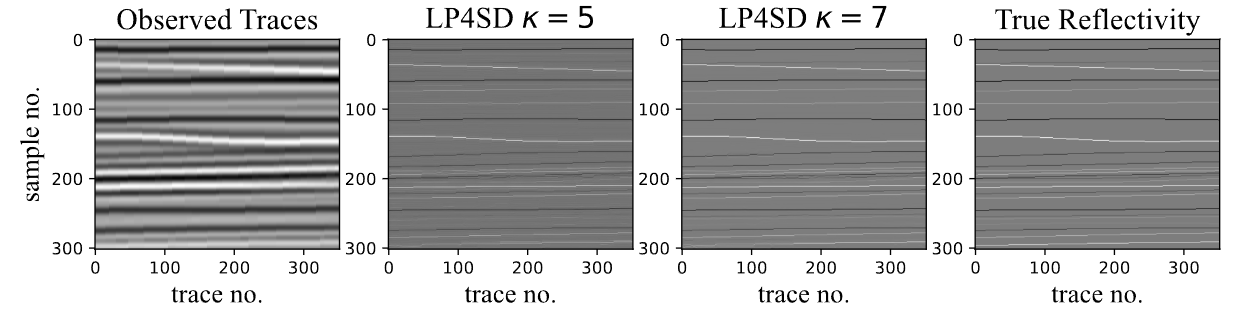}%
}
\caption{Evaluation results on 2D Marmousi2 data using LP4SD}
\label{fig:2Dimg_LU5_marm}
\end{figure}

While synthetic data is used to mimic the real trace and reflectivity, it is generally impossible to cover all possible cases in real distribution. However, it is observed that LP4SD generalizes better for reflectivities with unseen sparsity. Figure \ref{fig:1D_res_marmousi} compares the residuals of reconstruction $\boldsymbol{\hat{x}}-\boldsymbol{x}$ of the U-Net and LP4SD with kernel size 7 in different sparsity levels. When the sampling frequencies used during evaluation differ from those used while training (visualized/reflected as sparsity in the reflectivity), U-Net tends to produce more error. On the other hand, LP4SD is more robust to unseen sparsity. The numerical results are displayed in Table \ref{tab:1D_marmousi} and the sampled 2D results can be found in Figure \ref{fig:2Dimg_LU5_marm}. This is because LP4SD breaks down the inverse problem into $K$ proximal updates using the forward operator. The learning-free gradient step gives a better approximation of $\boldsymbol{x}$ by minimizing $||y-Ax||^2$, so that each proximal step (CNN) only learns a denoising process. Whereas U-Net learns the direct mapping from the seen distribution of seismic traces $\boldsymbol{y}$ to reflectivities $\boldsymbol{x}$, so it generalizes poorly when the trace is out of distribution, i.e., with unseen sparsity levels.

\begin{table}
    \centering
    \caption{\label{tab:1D_marmousi} Marmousi2 1D and 2D testing result, where the best metrics in each experiment is marked in bold.}
    \begin{tabular}{ ccccc } \hline 
         1D Marmousi2   & Methods    & MSE       & $\gamma$  &  Q \\
         \hline
                        & U-Net    & 0.002627      & 0.909      & 9.95  \\
         Full sample    & LP4SD ($\kappa=5$)   & \textbf{0.000318}      & 0.961      & \textbf{17.48} \\
                        & LP4SD ($\kappa=7$)   & 0.000390      & \textbf{0.963}      & 17.18  \\
        \hline
                        & U-Net     & 0.002214      & 0.916      & 9.52 \\
        Downsampled by 2 & LP4SD ($\kappa=5$)   & 0.000382      & 0.975      & 15.44   \\
                        & LP4SD ($\kappa=7$)    & \textbf{0.000334}      & \textbf{0.977}      & \textbf{16.29} \\
        \hline
                        & U-Net   & 0.002149      & 0.912      & 8.78 \\
        Downsampled by 3& LP4SD ($\kappa=5$)   & \textbf{0.000449}      & 0.975      & 14.10 \\
                        & LP4SD ($\kappa=7$)   & 0.000486      & \textbf{0.976}      & \textbf{15.07}    \\ 
        \hline \hline
        2D Marmousi2   & methods    & MSE       & $\gamma$  &  Q \\
         \hline 
                        & U-Net    & 0.001448      & 0.852      & 6.06  \\
         Full sample    & LP4SD ($\kappa=5$)   & \textbf{0.000053} & \textbf{0.991} & \textbf{18.80}  \\
                        & LP4SD ($\kappa=7$)   & 0.000369      & 0.977      & 15.01  \\
        \hline
                        & U-Net    & 0.002378      & 0.840      & 5.59\\
        Downsampled by 2& LP4SD ($\kappa=5$)   & 0.000709      & 0.952      & 11.67  \\
                        & LP4SD ($\kappa=7$)   &\textbf{0.000327}   & \textbf{0.975}   & \textbf{15.67}  \\
        \hline
                        & U-Net    & 0.002843      & 0.812      & 4.88 \\
        Downsampled by 3& LP4SD ($\kappa=5$)   & 0.000911      & 0.931      & 9.79 \\
                        & LP4SD ($\kappa=7$)   & \textbf{0.000137}     & \textbf{0.980}      & \textbf{16.74}  \\ \hline
    \end{tabular} 
\end{table} 

\subsection{Testing on Real Data}
\begin{figure}[htp]
\centering
\includegraphics[clip,width=0.6\columnwidth]{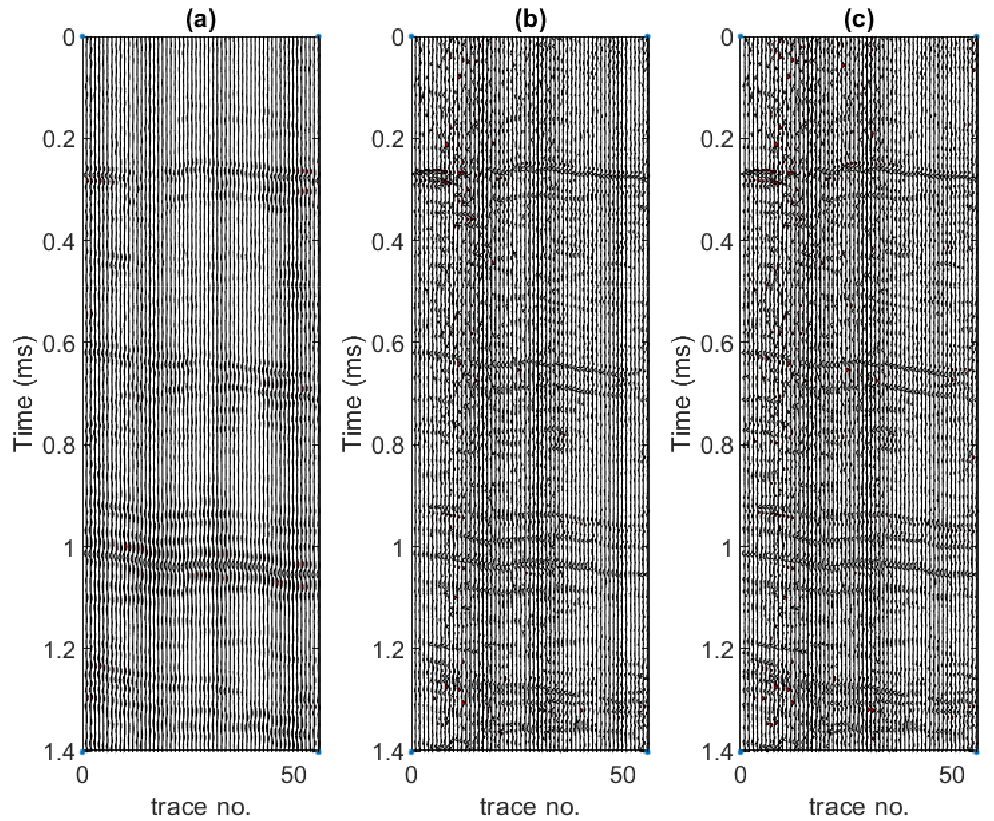}%
\caption{Evaluation results on 2D real data, (a) observed traces, (b) reconstructed reflectivity and (c) reconstructed reflectivity after AGC}
\label{fig:real_2DIMG}
\end{figure}

Finally, LP4SD is evaluated on the real dataset, where ground-truth reflectivity is unknown. A 2D landline from east Texas, USA \cite{mousa2011processing} is used in this work. The survey contains 18 shots, and 594 traces in line with each trace having 1501 samples and a time sampling interval of 2 ms. 
The Common mid-point (CMP) gather is extracted from raw data, which is then divided and padded into patches of size that can be fed to the network. 
Then Automatic Gain Control (AGC) is applied to correct the magnitude. Note that the missing traces are muted (filled with zeros).
Figure \ref{fig:real_2DIMG} (a), (b) and  (c) depict the observed traces, concatenated reconstructions obtained directly from the network output, and recovered reflectivity after applying AGC, respectively. The layers are emphasized and more structural details are revealed in the reconstruction.

\begin{figure*}[htp]
    \centering
    \includegraphics[width=\textwidth]{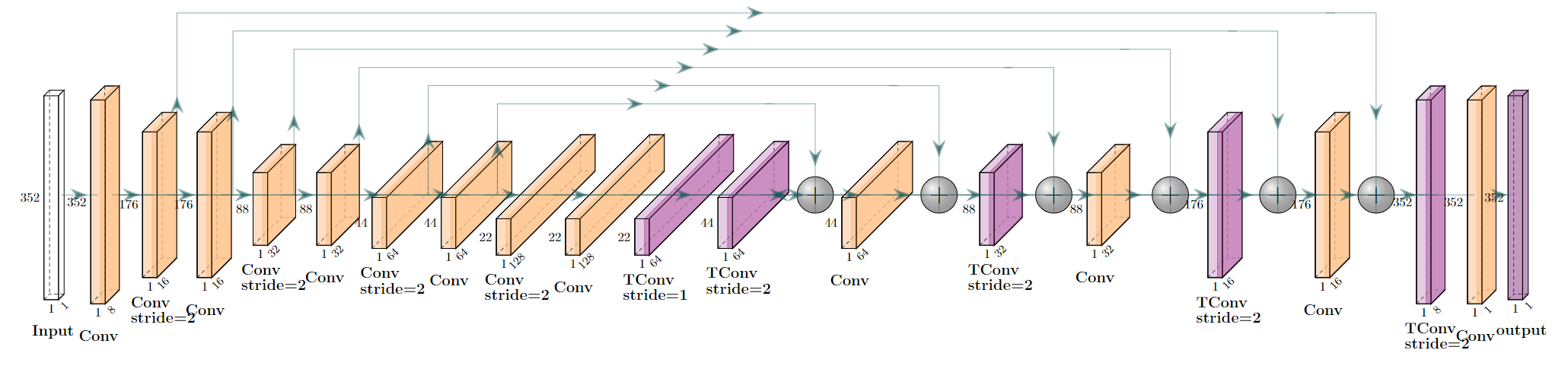}
    \caption{1D U-Net Architecture, normalization and activation layers are omitted in the figure for clarity.}
    \label{fig:AE_architecture}
\end{figure*}

\section{Conclusion} \label{sec:conclusion}
In this work, a Learned Proximal operator for Seismic Deconvolution (LP4SD) is proposed. The network architecture is both model-driven and data-driven. Hence, the proposed approach takes into account the strengths of model- and data-driven methods. LP4SD unfolded optimization iterations into a sequence of proximal gradient steps (model-driven) and replace the proximal operator with a 5-layer CNN (data-driven). The experimental test revealed the LP4SD's following advantages: 1) does not rely on predetermined assumptions on the reflectivity and thus avoids prior-specific parameter tuning, 2) breaks down the extremely challenging task of learning direct inverse into smaller manageable and easy-to-learn tasks of learning proximal operators, 3) depends on the forward operator thus more robust to unseen sparsity, and 4) can handle multiple traces (in 2D setup) simultaneously which is shown to give better reconstruction quality. In the experiments, the above advantages are observed and shown that LP4SD outperforms U-Net which learns the direct inverse in different noise levels.

\appendix 
\subsection*{Detailed Architecture of U-Net}
 Figure \ref{fig:AE_architecture} shows the U-Net for 1D input, whereas the U-Net for 2D input is similar to 1D, but adds two convolutional layers and two Transposed Convolutional layers in the middle of the network.

\bibliographystyle{unsrt}
\bibliography{reference.bib}

\end{document}